\documentclass[aps,prb,twocolumn,superscriptaddress,showpacs,english]{revtex4-1}

\usepackage[T1]{fontenc}
\usepackage[latin9]{inputenc}
\usepackage{babel}
\usepackage{amsmath}
\usepackage{amssymb}
\usepackage{wasysym}
\usepackage{graphicx}
\usepackage{xcolor}
\usepackage{bm} 

\newcommand{\UniOsaka}{Division of Materials Physics, Graduate School of Engineering Science, Osaka University, Osaka 560-8531, Japan}
\newcommand{\UniTokyo}{Department of Applied Physics, University of Tokyo, Tokyo 113-8656, Japan}
\newcommand{\Riken}{RIKEN Center for Emergent Matter Science, 2-1 Hirosawa, Wako, 351-0198, Japan}
\newcommand{\UniRoma}{Dipartimento di Fisica, Universit\`a di Roma La Sapienza, Piazzale Aldo Moro 5, I-00185 Roma, Italy}
\newcommand{\KEK}{Institute of Materials Structure Science, High Energy Accelerator Research Organization, Tsukuba, Ibaraki 305-0801, Japan}
\newcommand{\NIMS}{Research and Services Division of Materials Data and Integrated System (MaDIS), National Institute for Materials Science (NIMS), Ibaraki, Tsukuba 305-0047, Japan}
\newcommand{\UniOsakaCNRS}{Center for Spintronics Research Network (CSRN), Graduate School of Engineering Science, Osaka University, Osaka 560-8531, Japan}

\begin{document}

\title{High-pressure synthesis of Ba$_2$RhO$_4$, a rhodate analogue of the layered perovskite Sr-ruthenate}  

\author{I. Kurata} \affiliation{\UniOsaka}\affiliation{\UniTokyo}
\author{Jos\'e A. Flores-Livas} \affiliation{\UniRoma} 
\author{H. Sugimoto} \affiliation{\UniTokyo}
\author{H. Takahashi} \affiliation{\UniOsaka}\affiliation{\UniOsakaCNRS}
\author{H. Sagayama} \affiliation{\KEK}
\author{Y. Yamasaki} \affiliation{\NIMS}
\author{T. Nomoto} \affiliation{\UniTokyo}
\author{R. Arita} \affiliation{\UniTokyo}\affiliation{\Riken}
\author{S. Ishiwata} \affiliation{\UniOsaka}\affiliation{\UniOsakaCNRS}

\date{\today}

\begin{abstract}
A new layered perovskite-type oxide Ba$_2$RhO$_4$ was synthesized 
by a high-pressure technique with the support of convex-hull calculations.  
The crystal and electronic structure were studied by both experimental and computational tools. 
Structural refinements for powder x-ray diffraction data showed 
that Ba$_2$RhO$_4$ crystallizes in a K$_2$NiF$_4$-type structure, isostructural to Sr$_2$RuO$_4$ and Ba$_2$IrO$_4$. 
Magnetic, resistivity, and specific heat measurements for polycrystalline samples of Ba$_2$RhO$_4$ indicate that the system can be characterized as a correlated metal.  
Despite the close similarity to its Sr$_2$RuO$_4$ counterpart in the electronic specific heat coefficient and the Wilson ratio, Ba$_2$RhO$_4$ shows no signature of superconductivity down to 0.16\,K. 
Whereas the Fermi surface topology has reminiscent pieces of Sr$_2$RuO$_4$, an electron-like e$_{\rm{g}}$-($d_{x^2-y^2}$) band descends below the Fermi level, 
making of this compound unique also as a metallic counterpart of the spin-orbit-coupled Mott insulator Ba$_2$IrO$_4$. 
\end{abstract}

\maketitle

\section{Introduction}\label{Sec:Int}

Layered perovskite-type oxides A$_2$MO$_4$ (A: Rare-earth metal or alkaline earth metal; M: transition metal) have been extensively studied as two-dimensional correlated electron systems showing exotic electronic properties. This is typified by La$_{2-x}$Sr$_x$CuO$_4$ and Sr$_2$RuO$_4$ showing unconventional superconductivity~\cite{Cava1987-qp,Torrance1988-mk,Maeno1994-pg,Mackenzie2017-bz,Pustogow2019-el} and Sr$_2$CoO$_4$ showing ferromagnetic and metallic behavior~\cite{Matsuno2004-gk}. Because of the close analogy to La$_2$CuO$_4$, Sr$_2$IrO$_4$ has recently attracted great attentions.~\cite{Crawford1994-nk, SrIrO_Novel-J_PRL2008,RPB_2019_SrIrO4} The ground state of Sr$_2$IrO$_4$ has been described as a spin-orbit coupled Mott insulating state with antiferromagnetic ordering.~\cite{kim2009phase} 
In A$_2$MO$_4$, symmetry and local distortion of the M-O square lattice, which depend on the size of A-site ion, seem to play an important role for their electronic ground states. 
For instance, Sr$_2$RuO$_4$ has tetragonal symmetry with straight Ru-O-Ru bonds, whereas Ca$_2$RuO$_4$ known as an antiferromagnetic Mott insulator has large distortion of the Ru-O-Ru bonds~\cite{Fukazawa2000-op}. 
Recently, Ba$_2$IrO$_4$ has been synthesized by a high pressure technique and found to be tetragonal with straight Ir-O-Ir bonds, 
which are in contrast to those of Sr$_2$IrO$_4$ with orthorhombic distortion\cite{Okabe_BaIrO_PRB-2011}. 
Reflecting the absence of the lattice distortion, the magnetic ground state of Ba$_2$IrO$_4$ is free from the spontaneous magnetic moment caused by the Dzyaloshinskii-Moriya interaction, unlike the case of Sr$_2$IrO$_4$.

As mentioned above, Sr$_2$RuO$_4$ and Ba$_2$IrO$_4$ share the same tetragonal layered perovskite-type structure but show contrasting ground states. 
Therefore, it is tempting to explore novel electronic states in their relatives. 
Following this viewpoint, we take focus on the rhodium analogue, that is Sr$_2$RhO$_4$ with moderate electron correlation and spin-orbit coupling (SOC)~\cite{Itoh1995-bq,perry2006sr2rho4,Sandilands2017-bn,Haverkort2008-ud}. 
So far, solid solutions between Sr$_2$RhO$_4$ and the known compounds such as Sr$_2$Rh$_{1-x}$Ir$_x$O$_4$~\cite{Qi2012-nx} and Sr$_{2-x}$La$_x$RhO$_4$~\cite{kwon2019lifshitz} have been studied. 
However, these oxides have a lattice distortion unlike Sr$_2$RuO$_4$ and Ba$_2$IrO$_4$ with straight M-O-M bonds, which can be critical for the emergence of novel electronic states.  

In this work, we succeeded in the high-pressure synthesis of a new layered perovskite rhodate Ba$_2$RhO$_4$ and characterized its structural, magnetic, and electronic properties. 
From Rietveld refinements of the synchrotron x-ray diffraction data, Ba$_2$RhO$_4$ was found to crystallize in the tetragonal space group ($I$4/$mmm$), being isostructural with Sr$_2$RuO$_4$. 
In addition to this first member of Ruddlesden-Popper phases Ba$_{n+1}$Rh$_{n}$O$_{3n+1}$ with $n=$ 1, other possible formations of the other phases with $n=$ 2, 3, and $\infty$ (perovskite phase) under high pressures are also predicted in terms of their formation enthalpy. 
Although the signature of superconductivity has not been found at low temperatures down to 0.16\,K, the close similarity to Sr$_2$RuO$_4$ counterpart will be discussed. 

\begin{figure*}[t!]
\includegraphics[width=2.0\columnwidth,angle=0]{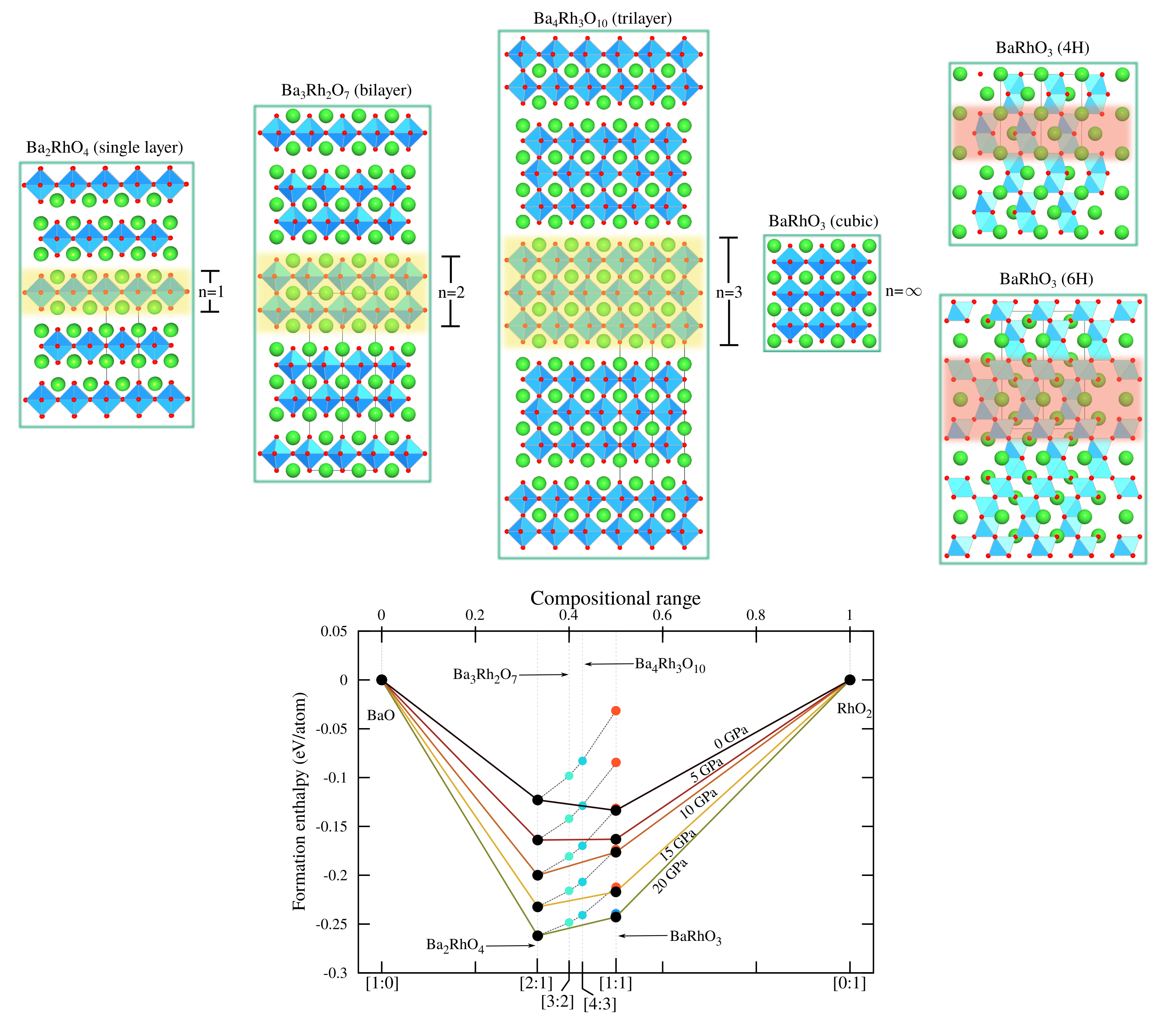}
\caption{~Bottom panel: Convex hull of enthalpy formation calculated for stoichiometries between BaO to RhO$_2$ as a function of pressure. 
Solid-black points represent stable compositions in the convex for selected pressures. 
For readability, the abscissa in the top show composition range and in the bottom composition ratio. 
Each of the points relates to a crystalline structure (show in the top) for different Ba-rhodates following the Ruddlesden-Popper sequence (Ba$_{n+1}$Rh$_{n}$O$_{3n+1}$).}
 \label{fig:Convex}
\end{figure*}

\section{Method}\label{Sec:Met}

\subsection{Experimental details}\label{Subsec:exp}

Polycrystalline samples of Ba$_2$RhO$_4$ were prepared by the solid-state reaction with a high-pressure technique using a cubic-anvil-type high-pressure apparatus (METEORITE).
The stoichiometric mixture of BaO$_2$(99.9$\%$, Furuuchi Chemical Co., Ltd.) and Rh metal powder (99.96$\%$, Furuya Metal Co., Ltd.) was encapsulated in a platinum tube in a glove box with a high-purity argon atmosphere. The sample was heated at 1350-1400 $^{\circ}$C for 30 min. at 8\,GPa, followed by quenching to room temperature before the pressure was released. 
The polycrystalline sample of Ba$_2$RhO$_4$ was obtained as a dense and black pellet.
As the sample is sensitive to air and moisture, it was kept in a glove box. The purity of the sample was checked by powder x-ray diffraction with Cu $K\alpha$ radiation (BRUKER new D8).
The data for structure refinements were collected as well, by synchrotron powder x-ray diffraction with a wavelength of 0.68975 \AA \quad at BL-8B, Photon Factory, KEK, Japan.
The diffraction data were analyzed using the Rietveld refinement using RIETAN-FP~\cite{Izumi2007-ni}. 
The resistivity and specific heat were measured by the standard dc four-probe method. Magnetic susceptibility data were collected by a superconducting quantum interference device magnetometer. 

\subsection{Computational details}

For the convex hull, all structural relaxations at different pressures were evaluated within DFT at the level of the Perdew-Burke-Erzernhof (PBE)~\cite{GGA-PBE} approximation to the exchange-correlation functional. 
A plane wave basis-set with a cutoff energy of 900\,eV was used to expand the wave-functions 
together with the projector augmented wave (PAW) method as implemented in the Vienna ab initio Simulation Package VASP~\cite{VASP_Kresse}. 
Structures in the convex hull construction were converged to less that two meV\AA $^{-1}$. 
All electronic band structures and magnetic properties were calculated using the full-potential 
linearized augmented plane-wave (FP-LAPW) method as implemented within the ELK code~\cite{elk-code,lejaeghere2016reproducibility}. 
We took into account spin-orbit-coupling (SOC) and performed a fully non-collinear LSDA magnetic calculation 
(searched for possible stable magnetic configurations, in case the system was magnetic). 
For coherence, in the last step, all volumes were optimized using the LDA functional).  
Finally, the potential and density were expanded in plane-waves with a cutoff of $| {\bf{G}}| =24/{a}_{0}$, 
and we set ${R}_{{\rm{\min }}}\times | {\bf{G}}+{\bf{k}}| $ to 9, where ${R}_{{\rm{\min }}}$ is the smallest muffin-tin radius. 
The maximum angular momentum $l$ for the expansion of the wave-function inside the muffin-tins set to 12. 
$k$-point grids were chose with a minimum allowed spacing between k points of 0.15 
in units of \AA$ ^{{-1}}$, equivalently to $14\times14\times8$ or more.  

\begin{figure}[t!]
\includegraphics[width=1.0\columnwidth,angle=0]{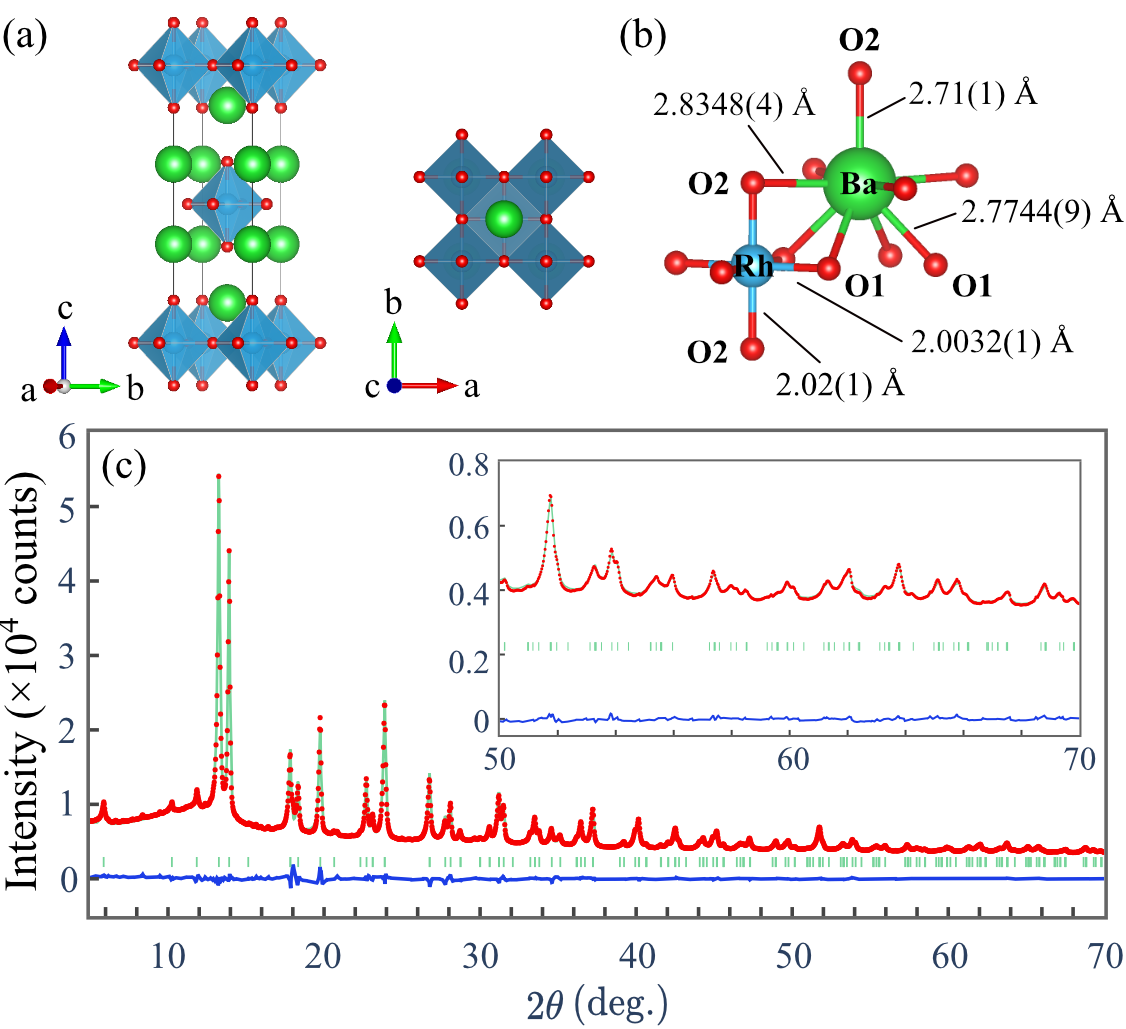}
\caption{~
(a) Crystal structure of Ba$_2$RhO$_4$ and 
(b) a schematic representation of the local structure with bond lengths around the Rh and Ba atoms. 
(c) Observed (red crosses), calculated (green line), and differential (blue line) synchrotron x-ray diffraction patterns. 
The green tick marks indicate the calculated peak positions.}
 \label{fig:Xray}
\end{figure}

\section{Results}\label{Sec:Res}

\subsection{Formation enthalpy of the 2-1-4 phase of Ba-Rh-O}\label{Subsec:enthalpy}

Fig.~\ref{fig:Convex} shows the convex hull of enthalpy formation calculated for stoichiometries between BaO--RhO$_2$, which can be found in the ternary phase diagram of Ba-Rh-O (see Fig. S1 in SM).
At ambient pressure, the formation energy suggests that compositions such as Ba$_2$RhO$_4$ (2-1-4) and BaRhO$_3$ (1-1-3) are likely to be accessible to synthesis. 
Under pressures above 5\,GPa, the 2-1-4 phase lowers its enthalpy, and is favorable with respect to the 1-1-3 phase (BaRhO$_3$), suggesting that could be stabilised upon the application of pressure. 
Our calculations also show that the Ruddlesden-Popper phases 
(Ba$_{n+1}$Rh$_{n}$O$_{3n+1}$) with bilayer ($n=$ 2) and trilayer ($n=$ 3) structures are not stable even at high pressures up to 20\,GPa. 
Completing the Ruddlesden-Popper phase with $n=\infty$, the enthalpy reaches the highest value and matches the 1-1-3 in its cubic-perovskite phase. 
It is worth mentioning that in our results, the lowest enthalpy structure for the 1-1-3 phase is disputed between the hexagonal 6H and 4H structure motif below 20\,GPa, 
and that only at very high pressures, the cubic perovskite phase dominates. 
Motivated by these findings, a high-pressure synthesis was carried out to find the stable 2-1-4 stoichiometry of the Ba-rhodate family (Sec. \ref{Subsec:exp}). 
We only attempted the synthesis of the 2-1-4 compositions, and after optimizing the synthesis conditions, structural and electrical measurements were carried out. 

\begin{table}[t!]
    \centering
    \begin{tabular}{cccccc}
        \hline \hline \text { Atoms } & \text { Site } & $x$ & $y$ & $z$ & $B$ (\AA$^{2}$) \\
        \hline Ba & 4e & 0 & 0 & 0.3556(1) & 0.60(2) \\
        Rh & 2a & 0 & 0 & 0 & 0.24(3) \\
        O1 & 4c & 0 & 0.5 & 0 & 1.0(1) \\
        O2 & 4e & 0 & 0 & 0.1522(7) & 1.0(1) \\
        \hline \hline
    \end{tabular}
    \caption{Structural parameters for Ba$_2$RhO$_4$. Space group $I4/mmm$ (No.139). $a=b=$ 4.0063(2) \ \AA, $c=$ 13.2966(8) \ \AA, $V=$213.42(2) \ \AA$^3$. Reliability factors: R$_{wp} =$ 2.486, S $=$ 1.95.}
    \label{table:rietveld}
\end{table}

\subsection{Structural properties}
Fig.~\ref{fig:Xray}(c) shows the powder x-ray diffraction pattern of Ba$_2$RhO$_4$, which was indexed on the basis of the tetragonal K$_2$NiF$_4$-type structure with space group $I4/mmm$ [$a = $ 4.0063(2) \AA, $c =$ 13.2966(8) \AA]. 
The model of the Rietveld analysis is summarized in Table~\ref{table:rietveld}, and the crystal structure is shown in Fig.~\ref{fig:Xray}(a). 
Another possible structural model with space group $I4_{1}/acd$, which allows rotation of the RhO$_6$ octahedra as in the case of Sr$_2$RhO$_4$, can be ruled out by the absence of the visible reflections which are allowed and forbidden for the models with $I4/mmm$ and $I4_{1}/acd$, respectively (see Fig. S2 in SM). 
Indeed, we confirmed that the reliability factors for $I4_{1}/acd$ are larger than for $I4/mmm$ (see Table SI). 
The Rh-O bond lengths 2.00$-$2.02\, \AA \, refined for $I4/mmm$ are fairly comparable to the value expected for that of Rh$^{4+}$$-$O$^{2-}$. 
On the other hand, as shown in Fig. 2(b), the refined Ba-O lengths are significantly shorter than the value 2.87\, \AA \ expected for the bond length between Ba$^{2+}$ (1.47\, \AA \ for the nine-fold coordination) and O$^{2-}$ ions (1.40\, \AA \ for the six-fold coordination)~\cite{shannon1970effective} indicating the compressed (over bonding) nature of the Ba-O bond. 
This feature points towards the metastable nature of the crystalline phase of Ba$_2$RhO$_4$ at ambient pressure. 


\begin{figure}[t!]
\includegraphics[width=0.9\columnwidth,angle=0]{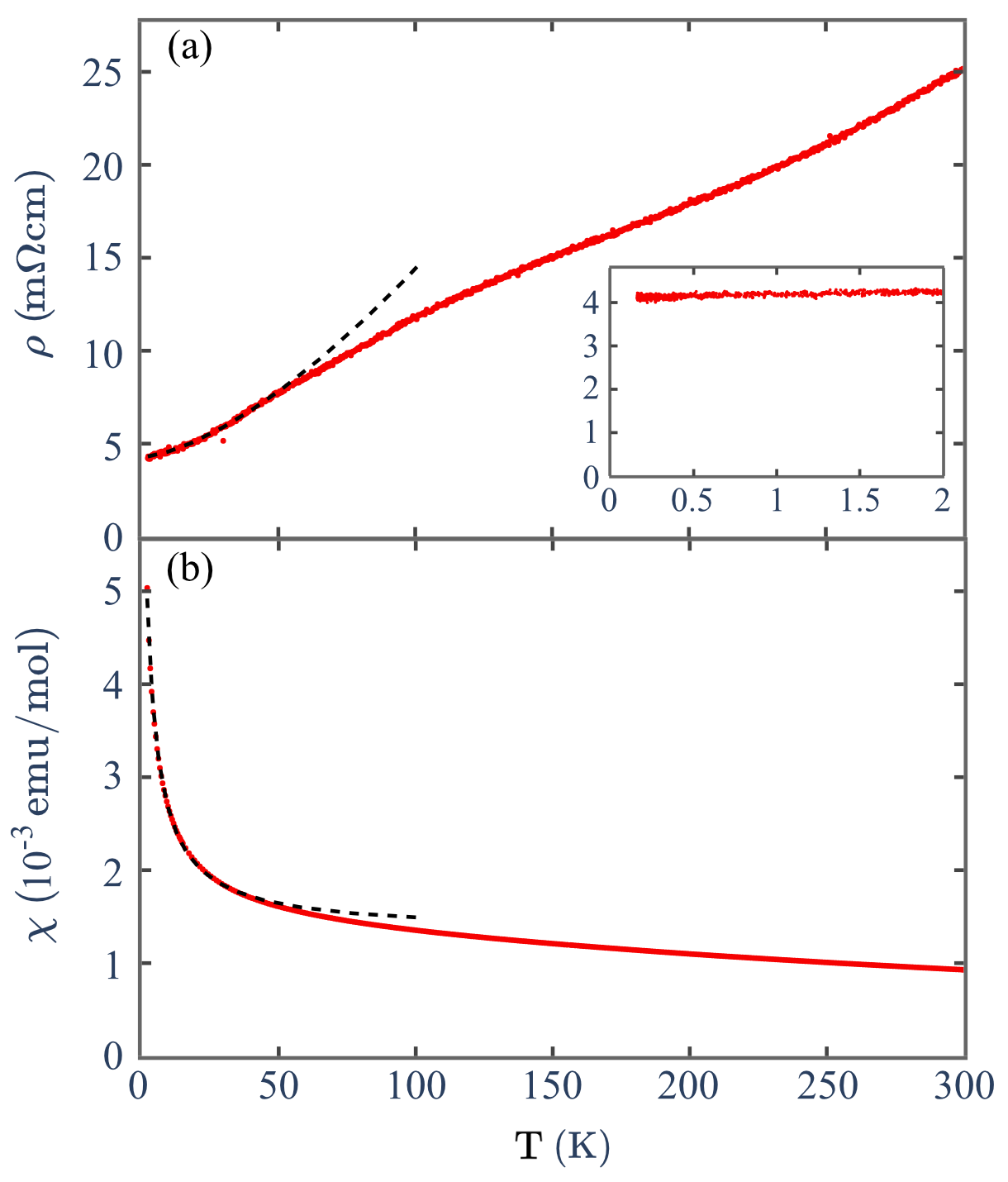}
\caption{~The temperature dependence of (a) electrical resistivity and (b) magnetic susceptibility for Ba$_2$RhO$_4$, respectively. 
The green dashed line shows the fitting of the experimental data with $\rho = \rho_0+\rm{A}$$T$$^{\alpha}$, and $\chi = \chi_0 + C/(T+\theta)$ below 50\,K, respectively. 
The inset of (a) shows no superconducting signals down to 0.16\,K.}
 \label{fig:rhochi}
\end{figure}

\subsection{Electrical resistivity, magnetic susceptibility, and specific heat capacity}

The electrical resistivity and magnetic susceptibility as a function of temperature for Ba$_2$RhO$_4$ are shown in Fig.~\ref{fig:rhochi}. 
Below 50\,K, the electrical resistivity deviates from the $T^2$ dependence, expected for a Fermi liquid, and resistivity evolves as 
$\rho = \rho_0+AT^{\alpha}$ with $\rho_0$=4.25 m$\Omega$cm, $A=$9.24\,$\mu\Omega$cm$/$K$^2$, and $\alpha$=1.52, which is in contrast to Sr$_2$RhO$_4$ with $\alpha$ comparable to or even larger than 2. 
Despite the apparent anomaly in $\alpha$ implying a non-Fermi liquid state, we cannot put in evidence its origin due to the polycrystalline nature of the material. 
It is also evident that we cannot conduct studies to rule out the possibility of anisotropy in the electrical resistivity inherent to the layered structure.
In fact, the residual resistivity $\rho_0$ is relatively large as reported for the polycrystalline sample of Sr$_2$RhO$_4$,~\cite{yamaura2004crystal} presumably reflecting the grain boundary scattering and the layered structure. 
Interestingly, even decreasing the temperature down to 160\,mK, no superconducting transition was observed [see the inset of Fig. 3(a)]. 
As shown in Fig. 3(b), a Curie-like tail is distinguishable below 50\,K, and fitting the magnetic susceptibility with the expression $\chi = \chi_0 + C/(T+\theta)$ gives $\chi_0 = 1.33\times 10^{-3} \ \text{emu/mol}$, $C = 1.63\times10^{-2} \ \text{emu/mol K}$, and $\theta = -2.55 \ \text{K}$.
The $C$ value is close to zero and no signal of magnetic transition is visible, indicating that a localized magnetic moment is absent in Ba$_2$RhO$_4$.

\begin{figure}[t!]
\includegraphics[width=1.0\columnwidth,angle=0]{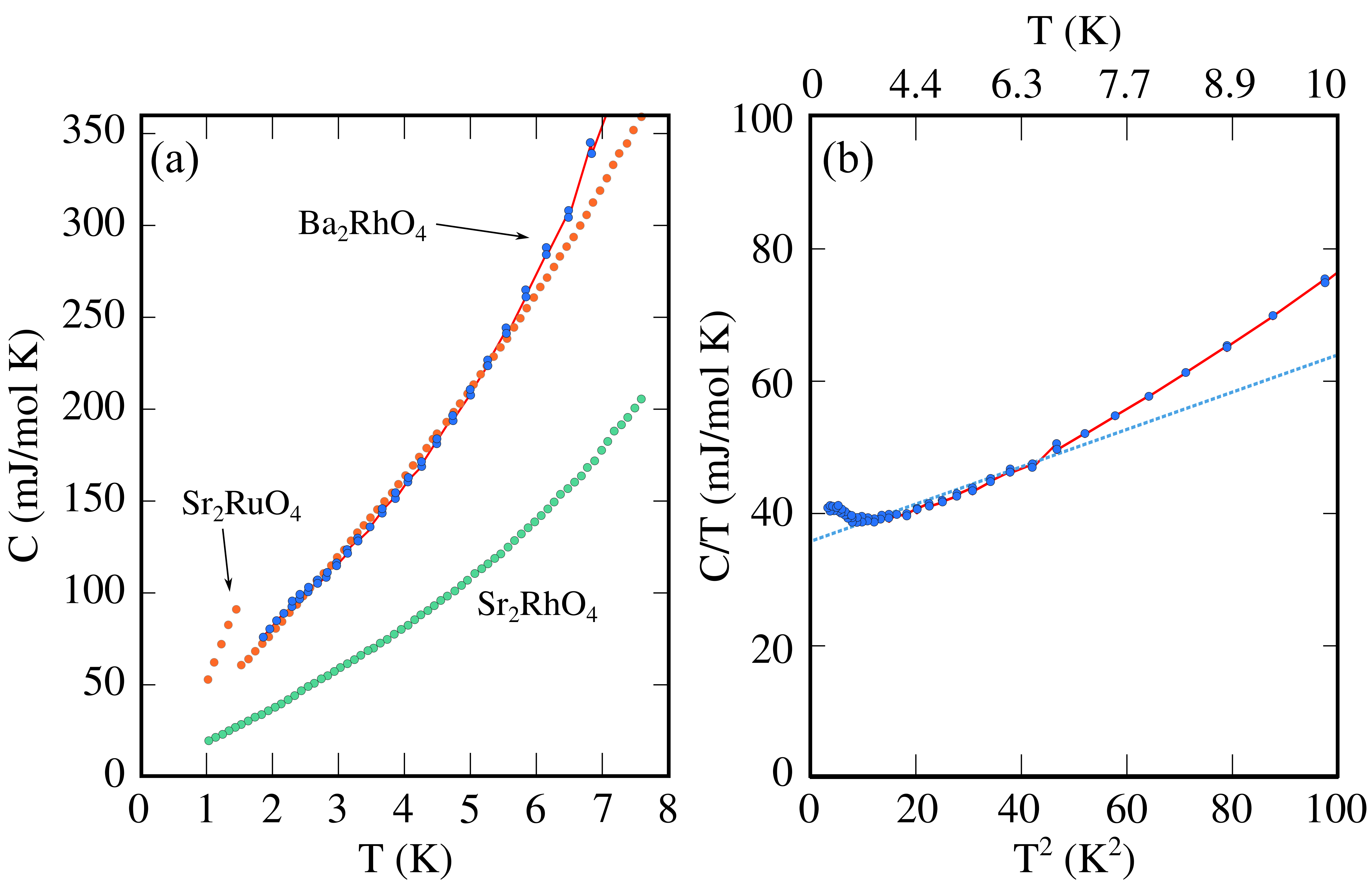}
\caption{
~(a): Temperature dependence of the specific heat for Ba$_{2}$RhO$_{4}$ (blue dots), the reported of Sr$_2$RuO$_4$ (red dots) and Sr$_2$RhO$_4$ (green dots). 
 (b): $T^2$ dependence of specific heat divided by temperature for Ba$_{2}$RhO$_{4}$ (blue dots). 
 The blue dashed line shows the fitting of the experimental data with $T/C=\gamma + \beta T^2$ between 2.0 and 7.5\,K.} \label{fig:Cv}
\end{figure}

The specific heat capacity of our Ba$_2$RhO$_4$ samples is shown in Fig.~\ref{fig:Cv}, and is compared to the reported ones of Sr$_2$RuO$_4$ and Sr$_2$RhO$_4$~\cite{perry2006sr2rho4,yamaura2004crystal}. 
The expected dominance of the linear electronic term is seen for both Ba$_2$RhO$_4$ and Sr$_2$RuO$_4$ below approximately 7\,K. 
Between 2.0 and 7.5\,K, the specific heat follows the equation $C / T=\gamma+(12 / 5) \pi^{4} N R \Theta_{\mathrm{D}}^{3} T^{2}(R=8.31 \mathrm{J} / \mathrm{mol} \mathrm{K} \text { and } N=1)$, where $\gamma$ and $\Theta_{\mathrm{D}}$ represent an electronic specific heat coefficient and Debye temperature, respectively. 
Using this relation, we obtained the values of $\gamma = 36.56$ \ mJ/mol K$^2$ and $\Theta_{\mathrm{D}} = 360$ K. 
The Debye temperature is comparable to that obtained by our first principle calculations ($\sim$ 320 K) and that of Sr$_2$RuO$_4$ ($\sim$ 410 K)\cite{Kittaka2018-jo}. To be noted here is that the electronic specific heat is as large as that of Sr$_2$RuO$_4$ ($\sim$ 39 \ mJ/mol)\cite{Maeno1994-pg}, implying that the electron correlation enhances the effective mass of conduction electrons. 
Here, we estimated the Wilson ratio $R_w$ using the following expression, 
\begin{equation*}
R_w = \dfrac{\pi^2 k_B^2}{3\mu_B^2}\dfrac{\chi_0}{\gamma}\, . 
\end{equation*}
The estimated $R_w(=2.64)$  is comparable to or even larger than the reported ones for Sr$_2$RuO$_4$ ($\sim1.9$) and Sr$_2$RhO$_4$ ($\sim2.3$), which are regarded as clean correlated electron metals~\cite{perry2006sr2rho4,georges2013strong}. 
One should notice that the specific heat follows the above equation only in the narrow temperature range. Although the origin of the deviation at low temperature is elusive, 
the deviations below 4\,K and above 7\,K may suggest an electronic instability that violates the current Debye model. 

\subsection{Band structure and Fermi surface}

\begin{figure}[t!]
\includegraphics[width=0.90\columnwidth,angle=0]{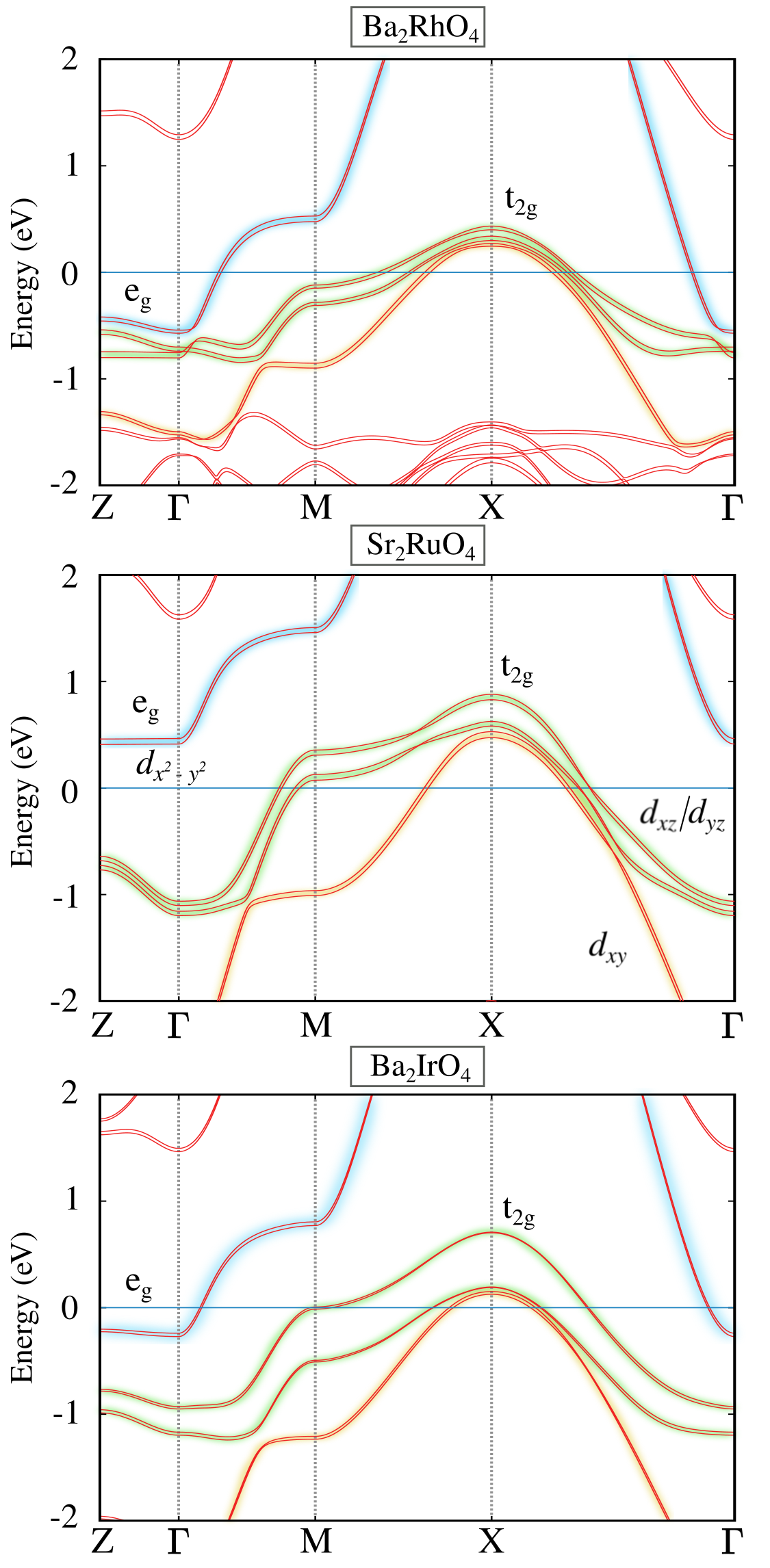}
\caption{~Full-potential DFT-LSDA band structure including spin-orbit coupling along selected high-symmetry points at zero pressure.}
 \label{fig:bands}
\end{figure}

Knowing that the 2-1-4 composition of Ba-Rh-O crystallizes in a layered-type structure similar to 
the 5$d$ counterpart Ba$_2$IrO$_4$, and the 4$d$, Sr$_2$RuO$_4$, it is interesting to compare the electronic band structure near the valence energy among these materials. 
The Kohn-Sham band structure of Ba$_2$RhO$_4$ calculated within the local density approximation (LSDA) and including SOC (LSDA+SO) is shown in Fig.~\ref{fig:bands}.  
To have a meaningful comparisons, we used their theoretical LDA-volume and assumed a K$_2$NiF$_4$-type structure for all cases.  
The band structure calculation, at the level of theory employed here, evidences a metallic and nonmagnetic character for Ba$_2$RhO$_4$ (see Table SII in SM), 
which is in agreement with our resistivity and magnetic measurements (see the calculated density of states presented in Fig. S3 in SM.). 
The case of Sr$_2$RuO$_4$, a prototypical system in manifolds, has been vastly characterised and our calculations agree with the reported band structure for this system~\cite{oguchi1995electronic,wang2017quasiparticle,PRX_Baumberger_2019,PRB_2020}. 
Ba$_2$IrO$_4$ is an interesting case on which has been reported basal-plane antiferromagnetism~\cite{Okabe_BaIrO_PRB-2011,origin_AF_BaIrO-NJP_2016,martins2017coulomb,arita2012ab,PRL_Silke_SrIrO_2011} 
at low temperatures and a Mott insulator transition is driven by the SOC~\cite{BS_BaIrO_2014_NJP}. 
While the level of theory to treat this system is beyond the current employed in this work, in a simplified picture, 
we wanted to compare with the energy bands on the 5$d$-counter part. 
The orbital character and electronic symmetry of the bands are depicted as follows; 
the t$_{2\rm{g}}$-dominated bands ($d_{xy}$, $d_{xz}$ and $d_{yz}$) are in yellow and green shade, and the e$_{\rm{g}}$ ($d_{x^2-y^2}$) bands are shown in blur-blue. 

As shown in Fig.~\ref{fig:bands}, the electronic structure of Ba$_2$RhO$_4$ is closely related to the one of Sr$_2$RuO$_4$, which has attracted a lot of attention for its possibility to emerge as a spin-triplet superconductor. 
However, the latest evidences suggest to disregard the chiral $p$-wave nature of the superconducting state of SrRuO$_4$~\cite{Pustogow2019-el,ishida2020reduction,Nat_Phys_SrRuO4_ultra-2020,Nat_Phys_SrRuO4_thermo-2020}. 
On the other hand, our new material, Ba$_2$RhO$_4$ has appealing differences evident around the $\Gamma$-point, 
where the e$_{\rm{g}}$ electron pocket descends the Fermi level, which is also found in Ba$_2$IrO$_4$ (using DFT-LSDA theory). 
A simple interpretation arises by looking at the position of the Fermi level in those systems; 
i.e., the electron counting of $d$ bands in Ba$_2$RhO$_4$ is expected to be one more than that in Sr$_2$RuO$_4$.
Looking at the t$_{2\rm{g}}$-dominated bands among the three compounds, and comparing to the NO-SOC bands (not shown) 
we can conclude that SOC plays an important role on describing the splitting of $d_{xz}/d_{yz}$ bands for the three compounds, 
while more marked differences arise for Sr$_2$RuO$_4$ and Ba$_2$IrO$_4$. 
In Ba$_2$IrO$_4$, the Mott insulator transition is driven by strong SOC, an effect that raises the bands opening a small band gap through a moderate Coulomb interaction. Although Ba$_2$RhO$_4$ is isoelectronic and isostructural with Ba$_2$IrO$_4$, such a gap opening is hampered by the reduced SOC. Thus, Ba$_2$Rh$_{1-x}$Ir$_{x}$O$_4$ with the tetragonal structure can be an ideal system to investigate the SOC-induced metal-insulator transition ~\cite{Qi2012-nx}.
As mentioned above, Ba$_2$RhO$_4$ is of peculiar interest, as it not only posses inherent bands from the Ru-counterpart, but also shares the similar electron bands from the 5$d$-Ir.  These feature places Ba-rhodate as strategic material that 
links the physics of 5$d$-oxides with 4$d$ with a metallic character~\cite{witczak2014correlated}. 

\begin{figure}[t!]
\includegraphics[width=1.0\columnwidth,angle=0]{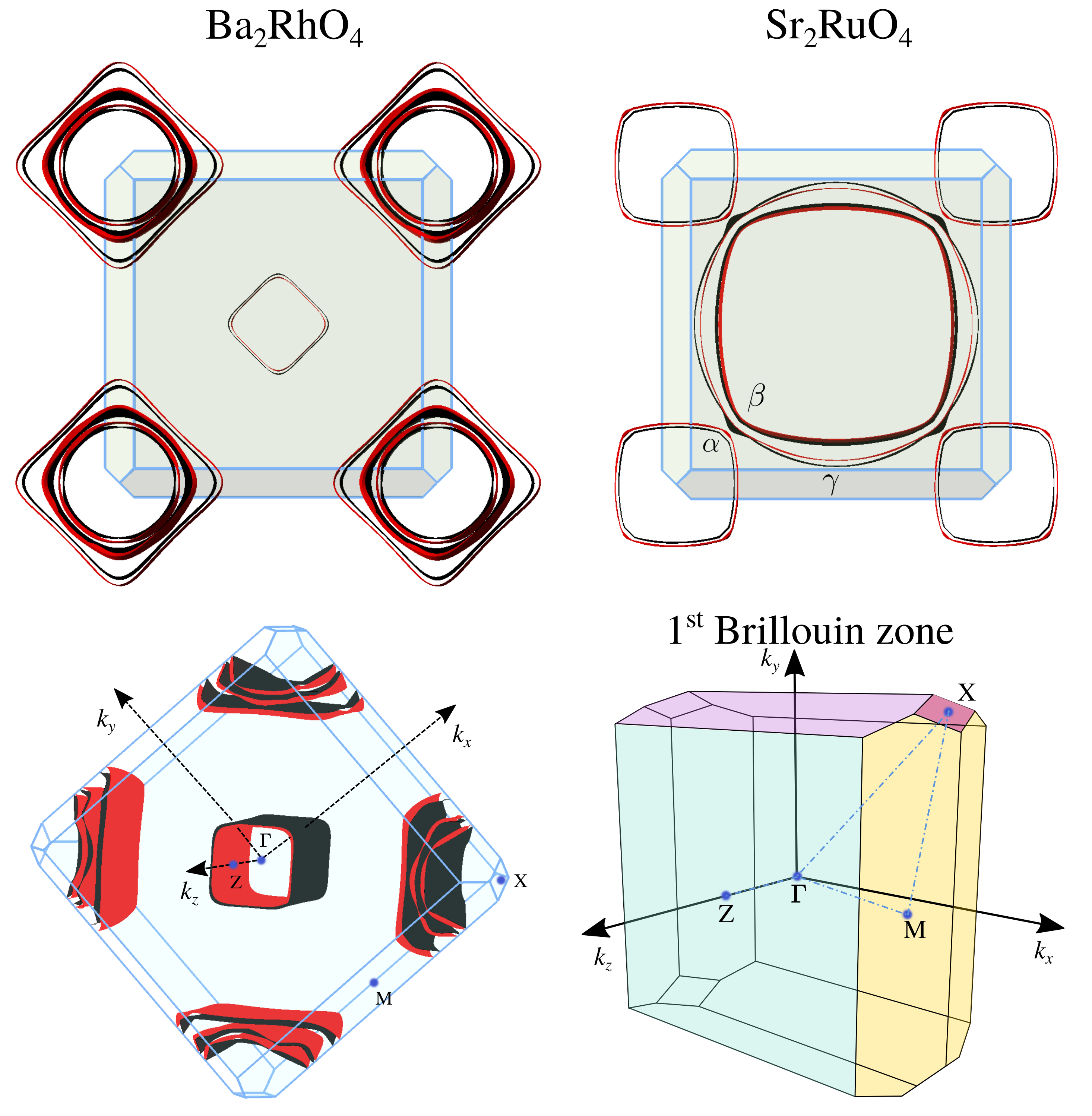}
\caption{~DFT-LSDA Fermi surfaces including spin-orbit coupling for Ba$_2$RhO$_4$ and Sr$_2$RuO$_4$.
two-dimensional projections are along $k_z$ vector. 
The first Brillouin zone and selected high-symmetry points for the structure with $I4/mmm$ are shown.}
 \label{fig:FermiSurface}
\end{figure}

Understanding the topology of the Fermi surface of Ba$_2$RhO$_4$ is important to unravel the nature of electrons in the normal phase and possibly to shed light on the superconducting mechanism. 
Top panels in Fig.~\ref{fig:FermiSurface} show two-dimensional Fermi surfaces of Ba$_2$RhO$_4$ and Sr$_2$RuO$_4$~\cite{wu2020electronic,PRL_nematic_2018}. 
Left bottom panel includes a three-dimensional cut, and in the right, the first-Brillouin zone for the $I4/mmm$-crystal as reference. 
The Fermi surface of Sr$_2$RuO$_4$ has been intensively investigated, both experimentally and theoretically\cite{oguchi1995electronic,wang2017quasiparticle,PRX_Baumberger_2019,PRB_2020}. 
{\it Ab initio} calculations based on the local-density approximation qualitatively reproduce the Fermi surface topology, provided that the SOC is taken into account. 
Our calculation reproduces the well-studied Fermi surface of Sr$_2$RuO$_4$, and also confirm that striking differences in the Fermi surface of Ba$_2$RhO$_4$ emerged. 
These difference are ascribed to the $d$-electron counting, and the electron pocket with e$_{\rm g}$ symmetry that pins below the Fermi level (see Fig.~\ref{fig:bands}). 
It should be noted that the square-like tubular sheets along with X-$\Gamma$ line can be regarded as quasi one-dimensional Fermi surface potentially yielding a nesting instability.
More recently, LDA+DMFT (local-density approximation + dynamical mean-field theory) calculations have emphasized the interplay of Coulomb interaction and t$_{2\rm{g}}$ crystal field and the role of the Hund's rule coupling~\cite{Malvestuto_PRB_2011,PRL_Pavarini-2016,PRL_Cote_SC-abinito}. 
Besides, laser-based angle-resolved photo-emission spectroscopy revealed the importance of SOC in Sr$_2$RuO$_4$, and the self-energies estimation~\cite{PRB_weak-coupling} 
for the $\beta$ and $\gamma$-sheets were found to display significant angular dependence~\cite{PRX_Baumberger_2019}. 
In this study, we did not perform calculations with higher theory beyond standard DFT, hence we cannot rule out that similar effects are present in Ba$_2$RhO$_4$.  
Finally, we also performed electron-phonon calculations for Ba$_2$RhO$_4$ using perturbation theory. 
The calculated phonon dispersion is shown in Fig. S4 in SM. The normal-state Eliashberg spectral function was adopted to estimate $\lambda$ (the dimensionless strength of the electron-phonon coupling) and the phonon average $\omega_{\rm {log}}$. 
These calculations reveal a weak BCS superconductor, indeed our estimation of T$_{\rm{c}}$ is well-below 0.1\,K, in agreement with the absence of any hint of superconductivity 
down to 0.16\,K in our experiments. 
To find the superconducting phase of Ba$_2$RhO$_4$, one might consider the possibility to further tune the electronic bands by applying pressure 
or by doing doping substitution, as recently reported for Sr$_{2-x}$La$_x$RhO$_4$ with a metal-insulator transition driven by a Lifshitz-transition~\cite{kwon2019lifshitz}. 
To have an insight into the pressure dependence of the bands, we also verified the dependence of the electronic structure under isotropic compression, from 1\,GPa to 20\,GPa. 
Ba$_2$RhO$_4$ remains metallic well above 20\,GPa, and the bands do not show any drastic change. 
The necessary pressure required to do a sizable change in the electronic structure is beyond 50\,GPa. However, phase decomposition and other enthalpically phases might become more favourable.
Another option to tune the bands with anisotropic pressure, i.e. due to the dimensionality of the layered perovskite phase, strain engineering via growth thins films is very attractive to search for the superconducting phase, proven that the synthetic route other than the current high pressure exists.


\section{Conclusions}\label{Sec:Con}
Using a high pressure technique, a new layered perovskite-type oxide Ba$_2$RhO$_4$ was synthesized and quenched to ambient conditions, which was supported by the convex hull calculations.  
The crystal and electronic structure were characterized by both experimental and theoretical works. 
Ba$_2$RhO$_4$ indeed shares many of the Sr$_2$RuO$_4$ with interesting properties; layered structure, specific heat, magnetic properties and overall electronic features. 
The magnetic and resistivity measurements indicate that the system can be characterized as a correlated metal. 
Interestingly, the material does not show signs of superconductivity down to 0.16\,K, which is supported by the theoretical estimation of the electron-phonon coupling. 
While the Fermi surface topology has reminiscent pieces of its Sr$_2$RuO$_4$ counterpart, one of the contrasting features in Ba$_2$RhO$_4$ is the presence of an electron e$_{\rm{g}}$ ($d_{x^2-y^2}$) band placed below the Fermi level, reflecting the increase in the $d$-electron counting. 
This feature provides a clue to discuss the role of electron correlation on the stabilization of SOC-driven Mott insulating state in the isoelectronic and isostructural counterpart Ba$_2$IrO$_4$. 
The new layered 2-1-4 perovskite phase, here reported for the first time~\cite{poole1999handbook}, adds new possibilities to characterize 4$d$-electron physics, and bridges to the 5$d$-electron that is driven by SOC effects to Mott insulators~\cite{NatPhys_Balents_2010}.  
The anomalous increase in the specific heat capacity below 4\,K and the absence of superconductivity will be subject of future work, conceding single crystals can be synthesized. 

\section{Acknowledgments}
Computational resources were provided by the Swiss National Supercomputing Center withing the project s970. 
This study was supported in part by KAKENHI (Grant No. 17H01195, 17H06137, 19H02424, 16H06345, 19H05825, and 19K14652), the Asahi Glass Foundation, and Multidisciplinary Research Laboratory System for Future Developments in Osaka University. 
The synchrotron XRD measurement was performed with the approval of the Photon Factory Program Advisory Committee (Proposal No.2018S2-006). We acknowledge the Center for Computational Materials Science, Institute for Materials Research, Tohoku University for the use of MASAMUNE-IMR (MAterials science Supercomputing system for Advanced MUlti-scale simulations towards NExt-generation - Institute for Materials Research), Project No. 20S0023.

\bibliographystyle{apsrev4-1}
\bibliography{bib}

\end{document}